\documentclass[9 pt,sigconf]{acmart}

\usepackage{amsmath}
\usepackage{microtype}
\usepackage{xcolor}
\usepackage{enumitem}

\copyrightyear{2026}
\acmYear{2026}
\setcopyright{cc}
\setcctype{by}
\acmConference[MobiHoc '26]{The Twenty-seventh International Symposium on Theory, Algorithmic Foundations, and Protocol Design for Mobile Networks and Mobile Computing}{November 23--26, 2026}{Tokyo, Japan}
\acmBooktitle{The Twenty-seventh International Symposium on Theory, Algorithmic Foundations, and Protocol Design for Mobile Networks and Mobile Computing (MobiHoc '26), November 23--26, 2026, Tokyo, Japan}
\acmDOI{10.1145/3842721.3850090}
\acmISBN{979-8-4007-2975-1/2026/11}

\begin{document}
\title{Demo: Closed-Loop Sionna--Isaac Sim Co-Simulation Framework for Wireless-Aware Robot Navigation over ROS~2}

\author{Yi Shen Lim, Seungeun Oh, Jihong Park}
\affiliation{%
  \institution{Singapore University of Technology and Design}
  \country{Singapore}
}
\email{{yishen\_lim, seungeun\_oh, jihong\_park}@sutd.edu.sg}

\renewcommand{\shortauthors}{Y. S. Lim et al.}

\begin{abstract}
A robot that offloads its control loop to the network carries the receiver with it, so link quality is decided by where it goes. Simulating this requires both a physics engine and a site-specific propagation model at once; to our knowledge no simulator natively unifies both, with existing couplings of the two limited to offline analyses. We demonstrate a real-time co-simulation framework coupling NVIDIA Isaac~Sim and NVIDIA Sionna over ROS~2 that closes the per\-cep\-tion--ac\-tion--com\-mu\-ni\-ca\-tion (PAC) loop between them. Sionna ray-traces the base-station-to-robot channel over the exact geometry Isaac~Sim simulates on---rather than modeling it stochastically---and feeds channel states back into the control loop in real time. Ray-traced on GPU, the coverage map is refreshed in ${\sim}16$\,ms (${\sim}60$\,Hz), fast enough for real-time control. To showcase the framework's utility, we implement a wireless-aware navigation application in an OpenStreetMap(OSM)-derived SUTD campus twin with two Nova~Carter robots: the closed-loop planner eliminates communication outage at only $+7.4\%$ traversal time over the shortest-path baseline (which spends $7.9$\,s of its $81.2$\,s run disconnected).%, validating that our framework successfully bridges millisecond-scale ray tracing with continuous robotic physics in a unified visualization environment.
\vspace{-6pt}
\end{abstract}




\maketitle
\vspace{-3pt}
\section{Introduction}
\label{sec:intro}

Cloud robotics, teleoperation, and autonomous logistics offload perception and decision-making to edge servers, so a robot's control loop depends on a low-latency wireless link.
Link quality varies with the robot's own position, as structures reflect, diffract, and block the signal. Where the robot goes will decide the connectivity it gets, coupling planning and communication into one PAC loop~\cite{rosnetsim}.
Safe deployment requires a platform where motion and site-specific propagation can be studied, before a robot meets a dead zone.

\begin{figure}[t]
    \centering
    \includegraphics[width=1\columnwidth]{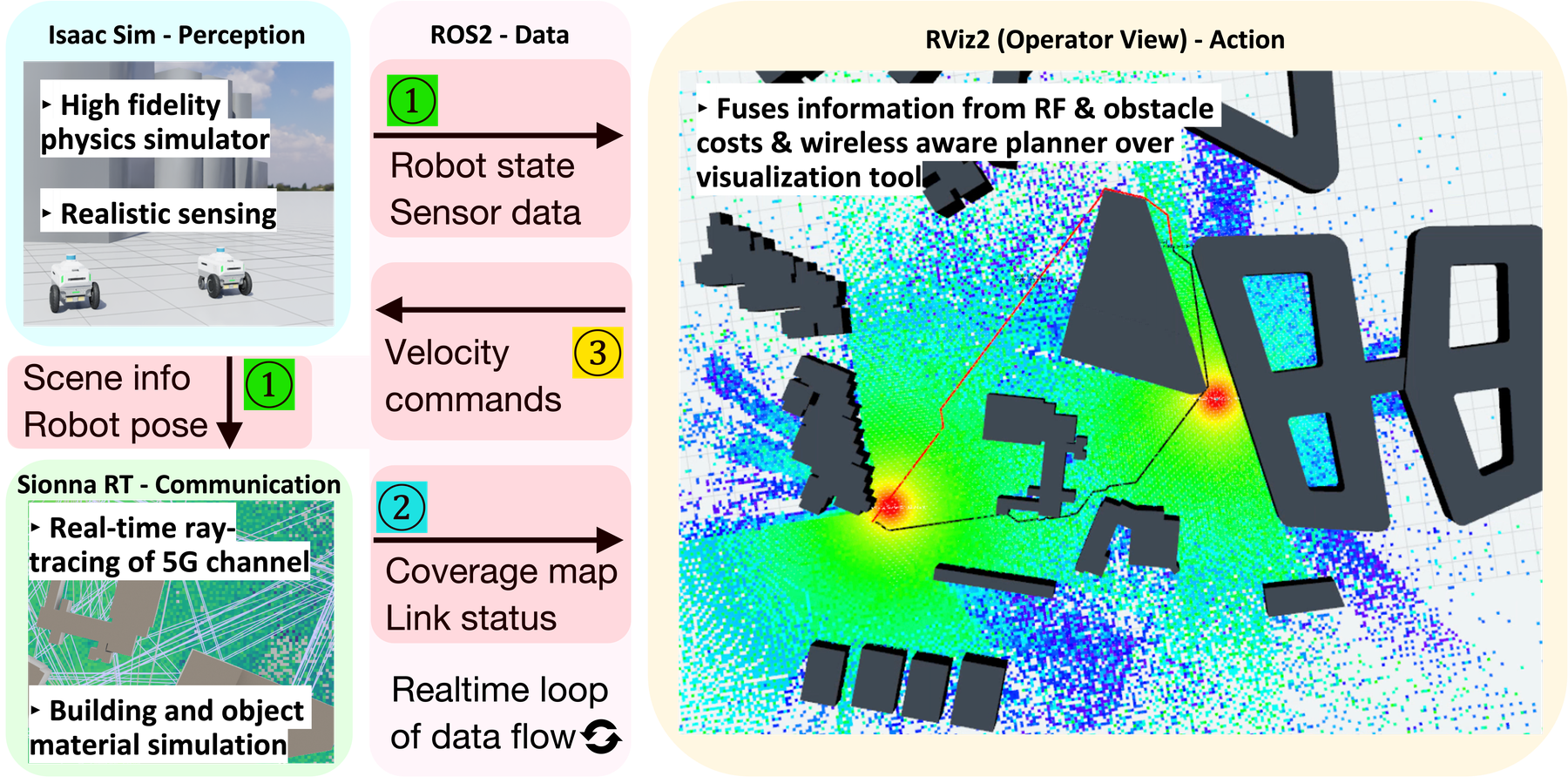}
\caption{Co-simulation view: Isaac~Sim simulates robots, Sionna ray-traces channels, ROS~2 relays states, RViz2 (ROS~2's visualization tool) overlays RSS heatmap and planned routes; attendees drive all views live in Scenario~1.}
\vspace{-3pt}
    \label{fig:arch}
        \vspace{-3pt}
\end{figure}

No single simulator offers this. Robotic simulators lack channel models, and wireless ones (Sionna~\cite{sionnart}, ns-3) lack physics.
Communication-aware planning~\cite{commrobotics} feeds the planner a statistical channel map; DT-RaDaR~\cite{dtradar} ray-traces an OSM/Blender twin with Sionna, as we do, but once and offline, with no physics engine.
Robot--network co-simulators~\cite{rosnetsim} close the PAC loop, but over stochastic propagation.
Ray-tracing twins such as CAVIAR~\cite{caviar}, AODT~\cite{aodt}, and SimART~\cite{simart} are site-specific but close the loop only on communication-side decisions (e.g., beam selection), with mobility given, not planned.
None, to our knowledge, combines site-specific ray tracing, a physics engine, and a channel that reaches the \emph{motion planner} online, at control-loop rates.

We demonstrate a framework that does, coupling \textbf{NVIDIA Isaac~Sim}~\cite{isaacsim} (a GPU-accelerated robotics simulator for physics and sensors) and \textbf{NVIDIA Sionna}~\cite{sionnart} (a differentiable wireless ray tracing library) over \textbf{ROS~2} (Robot Operating System~2, the standard robotics publish--subscribe middleware), inheriting planning (Nav2, ROS~2's navigation stack), robot interfacing, and visualization (RViz2) from the ROS~2 ecosystem.
Isaac~Sim, Sionna, and the planner exchange states over ROS~2 as asynchronous nodes. The ray-traced coverage map is published as a standard costmap layer that any ROS~2 planner can consume, and a campus-scale demo quantifies the resulting connectivity--travel-time trade-off.

The difficulty is not running the two simulators in parallel but \emph{unifying} them: both must share one scene geometry---so the ray-traced map registers cell-for-cell with the navigation grid and a moved obstacle casts physical and RF shadows simultaneously---while loops at milliseconds (ray tracing), tens of milliseconds (physics), and seconds (planning) must interlock without stalling.
We resolve the first with a single OSM-derived twin. Its Blender-extruded building footprints serve as Isaac~Sim's physics geometry, Sionna's ray-tracing scene, and the $0.5$\,m navigation grid, so ray-traced cells register with navigation cells and moved obstacles updates all three at once. We resolve the second in real-time, running asynchronous, message-triggered ROS~2 nodes (Sec.~\ref{sec:arch}).

\vspace{-6pt}
\section{System Architecture}
\label{sec:arch}

\vspace{-3pt}
\smallskip\noindent\textbf{(1) Perception: Scene, Physics \& State.}
We rebuild the $540{\times}540$\,m SUTD campus from OSM footprints via Blender, importing it into Isaac~Sim~\cite{isaacsim}, simulating two Nova~Carter robots at $3.0$\,m/s (Fig.~\ref{fig:arch}, left).
Poses (physics state) publish over ROS~2 at $60$\,Hz decoupled from the main loop (Fig.~\ref{fig:arch}, \textcircled{1}).
The $0.5$\,m navigation grid derives from the same twin; onboard sensing adds runtime obstacles.

\smallskip\noindent\textbf{(2) Communication: Ray Tracing \& Coverage.}
Control flows over the downlink as planning resides on the network. Two 5G base stations (gNodeBs) act as Txs delivering commands to the robot-mounted Rx, each at $18$\,dBm EIRP in band n78 ($3.425$\,GHz, $50$\,MHz), mirroring our physical private-5G small cell.
Both Txs sit $2.5$\,m on building facades (marked in Fig.~\ref{fig:arch}); the robot's vertically polarized dipole Rx (omnidirectional in azimuth) sits at $1.5$\,m.
Sionna RT~2.0.1's GPU ray tracer~\cite{sionnart} runs over the shared geometry ($2\times 10^6$ rays, $\le 5$ specular reflections) with ITU-R P.2040 concrete, brick, and metal across $48$ building meshes and a ground plane.
It computes RSS over a coarser $1.5$\,m grid of candidate Rx positions ($360{\times}360$ cells). The coverage map is sampled at the robot's live pose for instantaneous RSS; active Tx--Rx paths are rendered in Sionna's view (Fig.~\ref{fig:arch}, lower left).
Recomputing the map, triggered only by Tx motion or geometry changes, takes $\sim$16~ms (9.5--29.5~ms over 5 runs), i.e., ${\sim}60$\,Hz ($34$--$105$\,Hz), and publishes as a ROS~2 \texttt{OccupancyGrid} (Fig.~\ref{fig:arch}, \textcircled{2}).
\emph{Scope.} For scenes, channels are deterministic (RSS varies with position and changes only when the map is re-traced), poses are ground truth, channels are reciprocal, sharing one map, and association is best-server without inter-cell interference.

\smallskip\noindent\textbf{(3) Action: Wireless-Aware Motion Planning.}
To demonstrate how the framework closes the PAC loop, we instantiate a wireless-aware navigation application on top of its runtime ROS~2 topics.
The threshold $\mathit{RSS}_{\min}{=}{-}80$\,dBm is the lowest power at which our link budget ($7$\,dB noise figure, ${\approx}{-}90$\,dBm noise floor) sustains the command stream with margin.
Each cell takes a wireless cost $C_w = \min\bigl(254, \tfrac{254}{\Delta}\max(0, \mathit{RSS}_{\min}{-}\mathit{RSS})\bigr)$, with $\Delta{=}30$\,dB spanning Nav2's $[0,254]$ cost range (free at $\mathit{RSS}_{\min}$, lethal $\Delta$ below).
As a straightforward example to expose the PAC trade-off, we register this layer alongside physical obstacles $C_{\mathrm{obs}}$ (static map, detected obstacles, inflation) to give the combined traversal cost:
\vspace{-3pt}
\begin{equation}
    C_{\mathrm{total}}(x,y) = C_{\mathrm{obs}}(x,y) + \alpha \cdot C_w(x,y),
    \vspace{-3pt}
\end{equation}
which a custom A* planner searches (accumulating unit step costs per cell), though the framework's costmap interface can feed arbitrary downstream planners. Here, $\alpha{=}0$ recovers shortest-path planning and larger $\alpha$ buys detours around dead zones. A new search is triggered whenever the costmap changes, and the resulting velocity commands return to Isaac~Sim over ROS~2 (Fig.~1, \textcircled{3}), closing the loop that drives Scenario~1's interactive replanning.

\vspace{-6pt}
\section{Demonstration}
\label{sec:demo}

\vspace{-3pt}
\smallskip\noindent\textbf{Scenario~1: Interactive Closed-Loop Operation.}
Attendees set waypoints in the operator view (Fig.~\ref{fig:arch}) over the RSS heatmap; moving or adding gNodeBs and obstacles into the planned path recomputes the map and exposes handover boundaries.
Onboard sensing writes it into $C_{\mathrm{obs}}$, the re-traced map casts its RF shadow into $C_w$, and the robot detours around both a few seconds later.

\begin{figure}[t]
    \centering
    \includegraphics[width=1\columnwidth]{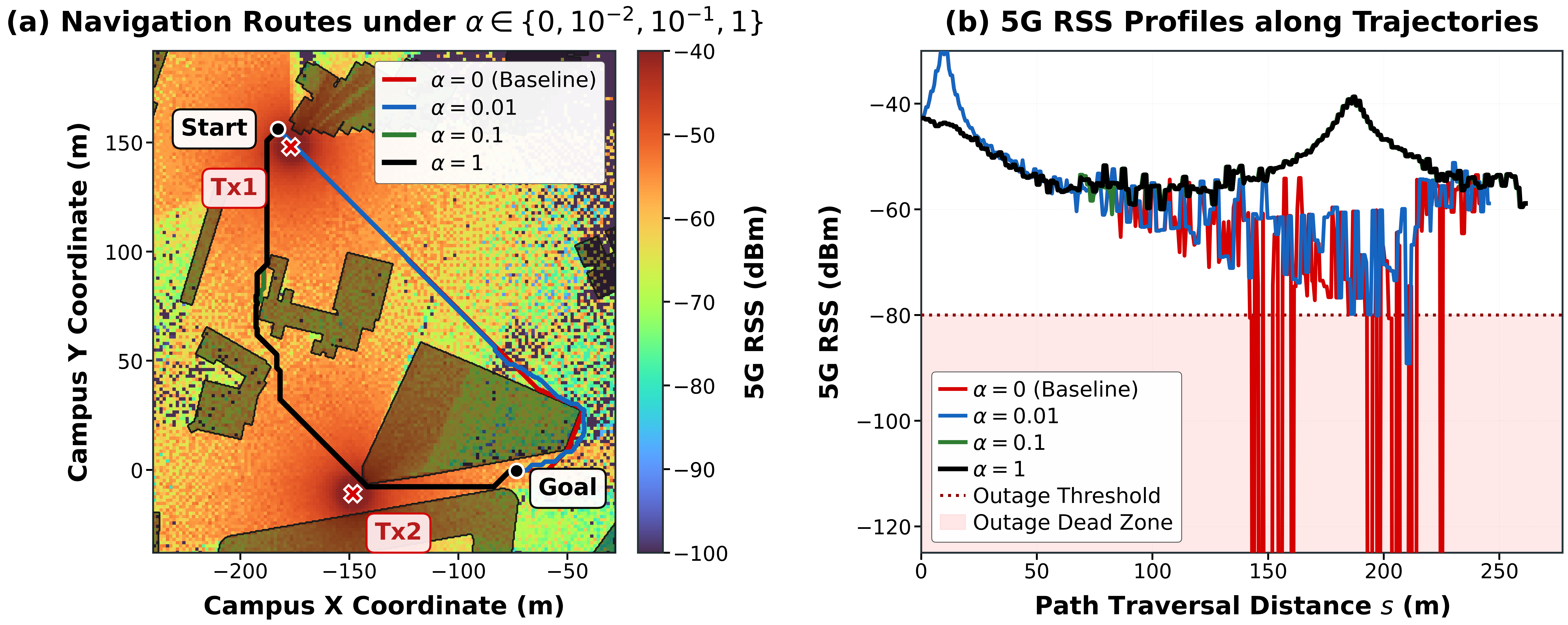}\caption{Routes (a) and RSS along each traversal (b) for $\alpha{=}0$ (red), $10^{-2}$ (blue), $10^{-1}$ (green), $1$ (black) over the ray-traced coverage map; Tx positions marked, sub-threshold region shaded; outage falls from $7.9$\,s (red) to $0$\,s (black).}
    \label{fig:comparison}
    \vspace{-6pt}
\end{figure}

\smallskip\noindent\textbf{Scenario~2: System Performance \& Closed-Loop Evaluation.}
\emph{System Latency ($10$ runs).} Within the steady-state loop, state ingestion ($0.83$\,ms), Sionna's ray-tracing kernel ($5.53{\pm}0.6$\,ms once warm; $118$\,ms initial JIT), and costmap fusion ($0.96$\,ms) deliver updated costmaps in $7.3$\,ms; the global A* search dominates, achieving a full loop at $4.73{\pm}0.10$\,s.
\emph{Closed-Loop Evaluation.} Each robot plans and traverses at $3$\,m/s in a static scene without halting on link loss, so every $\alpha \in \{0, 10^{-2}, 10^{-1}, 1\}$ yields a complete, comparable traversal (Fig.~\ref{fig:comparison}); \emph{outage} is time below $\mathit{RSS}_{\min}$.
The shortest-path baseline ($\alpha{=}0$) cuts through a building's RF shadow: $7.9$\,s of outage over an $81.2$\,s run.
$\alpha{=}10^{-2}$ leaves $1.3$\,s of outage at $+1.6\%$ traversal time, and by $\alpha{=}10^{-1}$ the route clears the shadow entirely ($87.0$\,s, $+7.4\%$, min.\ RSS $-60.9$\,dBm); larger $\alpha$ changes little.

\smallskip\noindent\textbf{Logistics.}
A single RTX~5090 workstation runs the three views side by side; attendees set goals in the GUI. We bring the workstation and two monitors and need one table (${\sim}1.5{\times}0.8$\,m), three outlets; no network, ${\sim}30$\,min setup. Video: \url{https://youtu.be/_TE7tGWxK6A}.

\vspace{-6pt}
\begin{acks}
This work was supported in part by the NVIDIA Academic Grant Program, in part by A$^\ast$STAR under its IAF-ICP (H25-MCP3438), in part by SUTD Kickstarter Initiative (SKI 2021 06 08), and in part by the National Research Foundation, Singapore and Infocomm Media Development Authority under its Communications and Connectivity Bridging Funding Initiative.
\end{acks}

\vspace{-6pt}
\bibliographystyle{ACM-Reference-Format}
\bibliography{main}

\end{document}